\documentclass[conference]{IEEEtran}
\IEEEoverridecommandlockouts

\usepackage{cite}
\usepackage{amsmath}
\usepackage{algorithmic}
\usepackage{graphicx}
\usepackage{textcomp}
\usepackage{xcolor}
\usepackage{tabularx,booktabs}
\usepackage{multirow}
\usepackage{multicol}
\usepackage{subcaption}
\usepackage{cleveref}
\usepackage{balance}
\usepackage{lastpage}
\usepackage{tabulary}
\usepackage{fancyhdr} 
\newcolumntype{C}{>{\centering\arraybackslash}X} 
\newcolumntype{L}{>{\raggedright\arraybackslash}X}

\def\BibTeX{{\rm B\kern-.05em{\sc i\kern-.025em b}\kern-.08em
    T\kern-.1667em\lower.7ex\hbox{E}\kern-.125emX}}
    
\graphicspath{{images/}}

\begin{document}

\title{Short-Term Load Forecasting for Smart Home Appliances with Sequence to Sequence Learning\\
}


\author{Mina Razghandi$^\dag$, Hao Zhou$^\ddag$, Melike Erol-Kantarci$^\ddag$, and Damla Turgut$^\dag$\\
{$^\dag$Department of Computer Science, University of Central Florida}\\
{$^\ddag$School of  Electrical Engineering and Computer Science, University of Ottawa }\\
mrazghandi@knights.ucf.edu, turgut@cs.ucf.edu, \{hzhou098, melike.erolkantarci\}@uottawa.ca \\


}

\renewcommand\floatpagefraction{.9}
\renewcommand\topfraction{.9}
\renewcommand\bottomfraction{.9}
\renewcommand\textfraction{.1}
\setcounter{totalnumber}{50}
\setcounter{topnumber}{50}
\setcounter{bottomnumber}{50}
\definecolor{cadmiumgreen}{rgb}{0.0, 0.42, 0.24}
\maketitle
\thispagestyle{fancy} %
      \lhead{} 
      \chead{Accepted by 2021 IEEE International Conference on Communications (ICC) , \copyright2021 IEEE } 
      \rhead{} 
      \lfoot{} 
      \cfoot{\thepage} 
      \rfoot{} 
      \renewcommand{\headrulewidth}{0pt} 
      \renewcommand{\footrulewidth}{0pt} 
\pagestyle{fancy}

\begin{abstract}
Appliance-level load forecasting plays a critical role in residential energy management, besides having significant importance for ancillary services performed by the utilities. In this paper, we propose to use an LSTM-based sequence-to-sequence (seq2seq) learning model that can capture the load profiles of appliances. We use a real dataset collected from four residential buildings and compare our proposed scheme with three other techniques, namely VARMA, Dilated One Dimensional Convolutional Neural Network, and an LSTM model. The results show that the proposed LSTM-based seq2seq model outperforms other techniques in terms of prediction error in most cases.




\end{abstract}





\section{Introduction}
\label{Introduction}

Appliance-level load forecasting for residential consumers is of significant importance for the modern power system, particularly considering the dynamic loads such as PV power and EVs that will stress the distribution system and consequently call for fine-grained control of micro-loads by home energy management systems\cite{Zenginis-2018-TII}. Therefore, mid-term and short-term load forecasting is essential for distribution system operation and control, while the long-term prediction plays a key role in infrastructure planning\cite{b8}. Short-term forecasting of household loads is closely related to consumer habits. An accurate and robust household load forecasting benefit the utility company and individual consumers. The utility company could better manage the electricity distribution and ancillary services while offering dynamic pricing to reduce peak demand. Meanwhile, individual consumers can better schedule the operation of their appliances to save from their electricity bills.

Electrical load forecasting is generally affected by many factors, e.g., such as weather conditions, social activity, consumers' habits, and so on. However, compared with load forecasting for a community or a commercial building, it is more challenging to forecast the short-term household load\cite{b5,b6,b12,b13,b14} due to the stochastic habits of consumers and the varying status and types of household appliances.

Several classification methods have been proposed in the literature for grouping similar load curves and eliminating the uncertainty\cite{b5}. The extraction of the consumption patterns of the electricity load can be accomplished by preprocessing the information using Fourier transforms or wavelet analysis\cite{b9}. The adoption of smart appliances by the consumers lead to changes to the traditional residential electrical load profiles\cite{b6}.

To this end, modern AI techniques, e.g., Recurrent Neural Networks (RNNs), Convolutional Neural Networks (CNNs), Deep Belief Network (DBN), have been employed for household load forecasting \cite{b12}. Neural networks are capable of extracting the nonlinear and non-stationary nature of energy consumption. For instance, CNN has been used for short-term load forecasting in\cite{b10,b13}, and the results show that CNNs outperform the Support Vector Machines (SVM), which is another machine learning algorithm. Furthermore, Long Short Term Memory (LSTM) network has been applied in\cite{b8,b9,b14,Razghandi-2020-Globecom} for load forecasting, which presents a lower prediction error compared with traditional methods. 


In this paper, we propose a novel LSTM-based Sequence to Sequence learning load forecasting method. While most previous studies mainly predict the overall household energy consumption, our work focuses on appliance-level short-term load forecasting, which has a higher level of uncertainty. 
Sequence to Sequence Learning (Seq2Seq) technique was initially proposed in\cite{sutskever2014sequence}, in which a multilayered LSTM network maps a sequence of input to a fixed-size vector and then, another multilayered LSTM network maps the fixed-sized vector to the sequence of target words. We use an LSTM network to map the sequence of past 24-hour energy consumption values to a fixed-size vector, then detect the appliance type, regenerate the input sequence in reverse form from the fixed-sized vector using another LSTM network, and produce a sequence of energy consumptions for the next hour from the fixed-length using another LSTM network. We show that Seq2Seq learning effectively deals with variable input and output sequence length and has more effective feature extraction abilities. We can see that our proposed model can distinguish the appliance type by its trend and learns appliances' typical usage duration. Our main contribution is the more accurate prediction of appliance-level consumption using a novel LSTM-based Seq2Seq learning algorithm.

The remainder of the paper is organized as follows. In Section II, we review previous work. We present our proposed  LSTM-based Sequence to Sequence learning load forecasting method in Section III. The evaluation study provides detailed results in Section IV. Finally, Section V concludes the paper.

\section{Related Work}
\label{RelatedWork}
There have been a number of studies in the electric load forecasting area, and the typical prediction methods used include i) traditional machine learning (e.g., Linear Regression, SVM, AutoRegressive Integrated Moving Average), ii) deep learning algorithms (e.g., RNNs, CNNs, and LSTM networks), iii) probabilistic forecasting (e.g., Quantile Regression, Density Regression), and iv) hybrid methods (e.g., combining CNNs with Gated Recurrent Units). 

Most recently, modern AI techniques, especially machine learning and deep learning, became the mainstream techniques for load forecasting\cite{b0}. A sparse coding-based learning approach for household electricity demand forecasting has been proposed by Yu et al.\cite{b1}, and 10\% higher accuracy has been observed compared to the existing studies. Based on multiple kernel learning, Wu et al.\cite{b2} proposed an improved gradient boosting framework, and the framework is extended to transfer learning context for both homogeneous and heterogeneous settings. Feng et al.\cite{b3} developed a two-step Q-learning, leveraging both deterministic and probabilistic load forecasting.

As stated earlier, recent years have witnessed the great potential of deep learning in many areas, which provides new opportunities for load forecasting as well. Bayesian deep learning has been applied in\cite{b4} to implement probabilistic load forecasting, and a clustering-based pooling method has been designed to prevent overfitting to improve the predictive performance. The overfitting problem has also been considered by Shi et al.\cite{b5}, where a novel pooling deep RNN is proposed to learn the uncertainty of load forecasting directly. Deep learning has also been used in non-intrusive load monitoring where no sub-metered information is needed to estimate the demand of individual appliances\cite{b6, b7}. 
LSTM network has been applied by Kong et al.\cite{b8} for short-term residential load forecasting and has been shown to outperform the conventional neural network-based approaches.

There are also a few hybrid methods proposed. For instance, LSTM has been combined with stationary wavelet transform technique for individual household load forecasting in\cite{b9}. CNNs and Gated Recurrent Units have been unified in\cite{b10}, resulted in lower computational complexity and higher prediction accuracy.

The inputs and targets should be encoded with vectors of fixed dimensionality to apply the deep neural network, which is a significant limitation. Different than prior works, in this paper, the sequential problem is solved by sequence to sequence learning. The idea of sequence to sequence learning is to use a multilayered LSTM to map the input sequence to a vector of a fixed dimensionality, and then another deep LSTM to decode the target sequence.

\section{Load Forecasting with Sequence to Sequence Learning}
\label{ProposedModel}

In this paper, we use a Sequence to Sequence (Seq2Seq) encoder/decoder model to predict the energy consumption of smart home appliances, which represents the behavioral pattern of smart home residents. The Seq2Seq encoder/decoder model derives occupants' habits from historical data and predicts hour-ahead appliance-level energy consumption in a household.

Sustkever et al.\cite{sutskever2014sequence} first introduced the sequence to sequence learning technique in 2014. The idea was to encode a sequence of inputs to a fixed-length vector with a multilayered LSTM and then decode the fixed-length vector to a sequence of outputs with another multilayered , noting that input and output sequences length can vary. They used Seq2Seq learning to translate sentences from English to French.

\begin{figure*}
    \centering
        \centering
        \includegraphics[width=0.75\textwidth]{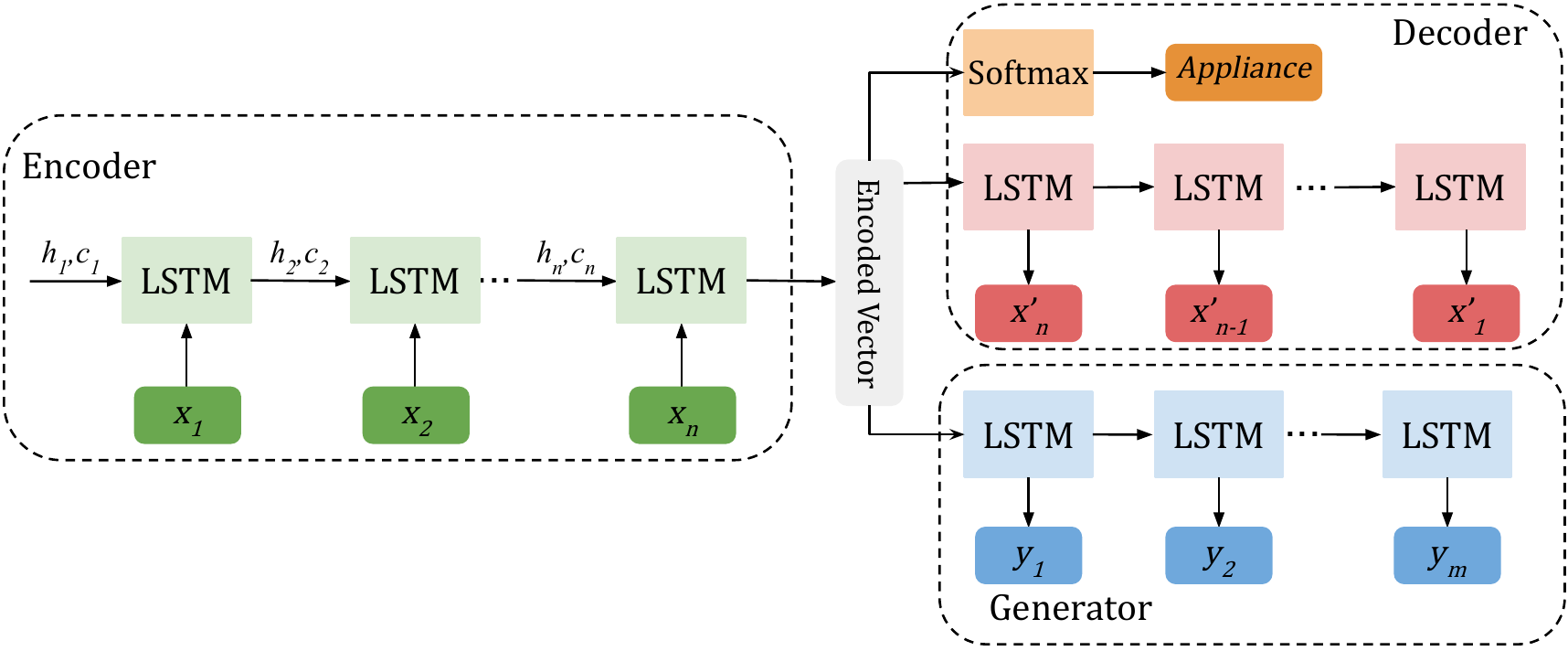}
        \caption{Seq2Seq encoder decoder model.}
        \label{fig:model}
        \vspace{-5mm}
\end{figure*}

We employed Seq2Seq learning to predict the energy consumption of smart home appliances for the next hour based on the previous 24 hours of historical data. As shown in Fig.\ref{fig:model}, our network architecture has three main LSTM network modules.

Considering $(x_{1}, x_{2},...,x_{n})$ as the input sequence for the encoder module, the LSTM \textbf{encoder} network maps the input sequence to a fixed-length vector $Z$, which is mainly based on the last hidden state of LSTM. The initial hidden state of the LSTM network in the \textbf{decoder} module is set to the fixed-length vector $Z$ and its goal is to compute and maximize the conditional probability of sequence $(x_{n}^{'}, x_{n-1}^{'},...,x_{1}^{'})$ of $(x_{1}, x_{2},...,x_{n})$:

\begin{equation}
\small
    p\left ( x_{n}^{'},...,x_{1}^{'} \vert x_{1},...,x_{n}\right)=\prod_{t=1}^{n}p\left ( x_{n-t}^{'}\vert x_{n}^{'},...,x_{n-t+1}^{'},Z \right)
    \label{eq:decoderprob}
\end{equation}

In the Equation~\ref{eq:decoderprob}, $(x_{n}^{'}, x_{n-1}^{'},...,x_{1}^{'})$ is the reversed form of input sequence $(x_{1}, x_{2},...,x_{n})$ enforcing encoder network to obtain long term dependencies of input sequence with putting more weight on most recent ones by minimizing a weighted MSE loss function (Equation~\ref{eq:decoderloss}). Besides this, by applying a Softmax function to the fixed-length vector $Z$, the decoder module learns the appliance working pattern and
predicts the type of appliance that was not given in the input features. 

\begin{equation}
    Weighted MSE = \frac{1}{n}\sum_{i=1}^{n}W * (x_{i}^{'} - x_{i})^{2}
    \label{eq:decoderloss}
\end{equation}

Finally, the LSTM network in the \textbf{generator} module, estimates the conditional probability $p\left ( y_{1},...,y_{m} \vert x_{1},...,x_{n}\right)$ according to the Equation~\ref{eq:genprob}:

\begin{equation}
    p\left ( y_{1},...,y_{m} \vert x_{1},...,x_{n}\right)=\prod_{t=1}^{m}p\left ( y_{t}\vert y_{1},...,y_{t-1},Z \right)
    \label{eq:genprob}
\end{equation}

In Equation~\ref{eq:genprob}, $(y_{1},...,y_{m})$ is the prediction results for next $m$ data points obtained from the fixed-length vector $Z$ by minimizing a MSE loss function:

\begin{equation}
    MSE = \frac{1}{m}\sum_{i=1}^{m}(y_{i} - \hat{y_{i}})^{2}
    \label{eq:genloss}
\end{equation}

\section{Evaluation Study}
\label{Experiments}

\subsection{Experiment Setup}
\label{Experimentsetup}
In the evaluation study, we used a publicly available dataset, GreenD\cite{monacchi2014greend}, collected from eight different households, including apartments and detached houses, from December 2013 to October 2014 in Austria and Italy. After data-processing, only four buildings had sufficient data for training a deep learning model (more than 150 days) with $\frac{1}{60}$ Hz frequency. This dataset contains appliance-level energy consumption for an average of nine appliances per household at 1Hz frequency. However, not all of the monitored appliances display a repeated pattern of the occupant's lifestyle. Therefore, Table~\ref{tab:datainfo} presents the five selected home appliances that were in common among all four buildings and had a repetitive trend. With 70\%, 20\%, and 10\% ratio, we split data into three training, validation, and test sets.

\begin{table}
\centering
\caption{Datasets used.}
\label{tab:datainfo}
\begin{tabular}{p{1.75cm}p{0.8cm}p{3.7cm}}
\toprule
\textbf{Building No.} & \textbf{Days} & \textbf{Monitored Appliances}                                     \\ \midrule
Building 0                & 236                     & dishwasher, lamp, fridge, radio, washing machine   \\
\hline
Building 1                & 226                    & dishwasher, lamp, fridge, radio, washing machine   \\
\hline
Building 2                & 231                     & dishwasher, TV, network attached storage, laptop, washing machine \\
\hline
Building 3                & 151                     & dishwasher, TV, fridge, computer, washing machine                 \\ \bottomrule
\end{tabular}
\vspace{-3mm}
\end{table}

We used three load forecasting methods from the literature as the baseline for evaluating the performance of the proposed model:
\begin{itemize}
    \item \textbf{Vector Auto Regression Moving Average (VARMA)}: This model is a combination of Vector Auto-Regression (VAR) and Vector Moving Average (VMA) models. VAR is a generalization of the Auto-Regressive model and used when there are multiple parallel variables to forecast. VAR models the variables as a linear function of their past variables, and the order of the VAR model determines the number of earlier time steps the model utilizes for predictions. VMA predicts the next steps based on a linear function of residual errors, that is, the difference between predicted and observed values to predict.
    \item \textbf{Dilated One Dimensional  Convolutional (CONV1D)}: This model is based on the WaveNet network, proposed in\cite{oord2016wavenet} that has dilated causal convolution layers to transform text to speech. The 1-dimensional convolution slides a filter on an input series usually by one stride, but then in the causal convolution, the ordering of the input data will remain the same, and the model will not be learning based on future data. In a dilated convolution, the sliding filter skips input data with certain steps. Multiple stacked dilated convolutional layers allow larger input sequences but keep the network complexity efficient.
    \item \textbf{Long Short-Term Memory Neural Network}: LSTM network is a type of Recurrent Neural Network specifically designed to learn long-term dependencies in input data by relying on feedback from previous stage outputs. We used a 2-layer LSTM network in this paper. 
    \item \textbf{Seq2Seq Learning}: As described in Section~\ref{ProposedModel}.
\end{itemize}

\subsection{Performance Indices}
\label{performanceIndices}
We have used the Root Mean Square Error (RMSE) and Mean Absolute Error (MAE) metrics to evaluate regression prediction accuracy. However, since both RMSE and MAE are scale dependant, we have also considered Normalized Root Mean Square Error (NRMSE) in our experiments. RMSE, MAE and NRMSE are shown in Equations~\ref{eq:rmse}--\ref{eq:nrmse} where $\mathit{n}$ is the total number of samples, $\mathit{y_{j}}$ is the target value and $\mathit{\hat{y_{j}}}$ is the predicted value, $\mathit{max(y_{j})}$ and $\mathit{min(y_{j})}$ refer to the maximum and minimum electrical usage recorded for the appliance $\mathit{j}$ in the data, respectively.
\begin{equation}
  RMSE = \sqrt{(\frac{1}{n})\sum_{i=1}^{n}(y_{j} - \hat{y_{j}})^{2}}
  \label{eq:rmse}
\end{equation}
\begin{equation}
  MAE = \frac{1}{n}\sum_{i=1}^{n} \vert y_{j} - \hat{y_{j}\vert}
  \label{eq:mae}
\end{equation}
\begin{equation}
  NRMSE = \frac{RMSE}{max(y_{j})-min(y_{j})}
  \label{eq:nrmse}
\end{equation}

\subsection{Data Preprocessing}
\label{datapreprocessing}

Linear interpolation is used to complete missing data. Moreover, instead of using minute-by-minute energy consumption of appliances, we smooth out the energy consumption of appliances for $T$=10 minutes without losing generality. In our dataset, appliance energy consumption varies between 10W to 2000W; therefore, the values are normalized before used in deep learning models.

Trends in energy consumption for electrical appliances can expose the living habits of household inhabitants. Some appliances would be used daily, such as TV or refrigerators, while others in a weekly fashion, such as vacuum cleaner, and some of them display no specific usage patterns.
It is worth noting that appliances with repetitive usability patterns are more suitable candidates to make predictions based on them. Accordingly, the historical data of power consumption and the day of the week are selected as the network input features. The row index decides the minute and hour specification when the dataset is sorted in terms of time, so they are no longer needed as an additional feature input. Instead, for this model, the input sequence length of 144 points is chosen with $T$-minute intervals covering 24 hours.



\subsection{Evaluation Results}
\label{Results}

Table~\ref{tab:EV-results} summarises RMSE, NRMSE, and MAE findings of four load forecasting algorithms for Building 0-3 with the lowest error in bold. Figure~\ref{fig:res}\subref{fig:dishwasher}--\ref{fig:res}\subref{fig:tv} plot and compare the forecasting performance of all four prediction algorithms in a day snapshot against the ground truth for dishwasher, lamp, fridge, radio, laptop and TV. 

The results demonstrate that the LSTM-based Seq2Seq and LSTM model, respectively, outperform the other forecasting models and have noticeably lower prediction error. LSTM model architecture is quite similar to the Seq2Seq model architecture (encoder, decoder, and generator). Consequently, the LSTM model has the closest results to the Seq2Seq model; however, the Seq2Seq model does a better job of learning the appliance's working pattern and detecting the appliance type. Unlike the LSTM model, the Seq2Seq model has no information about appliance type as an input.

As shown in Figure 2c, all four prediction models were effective in learning patterns for the appliance with the most seasonal energy usage pattern (fridge), which means that these models are more promising when there is a recurring pattern of behaviors of smart home residents. However, the Seq2Seq model performs better in load forecasting for appliances with a higher level of uncertainty in usage patterns.


\begin{figure*}
    \centering
    \vspace{-5mm}
    \begin{subfigure}{0.7\textwidth}
        \includegraphics[width=\linewidth]{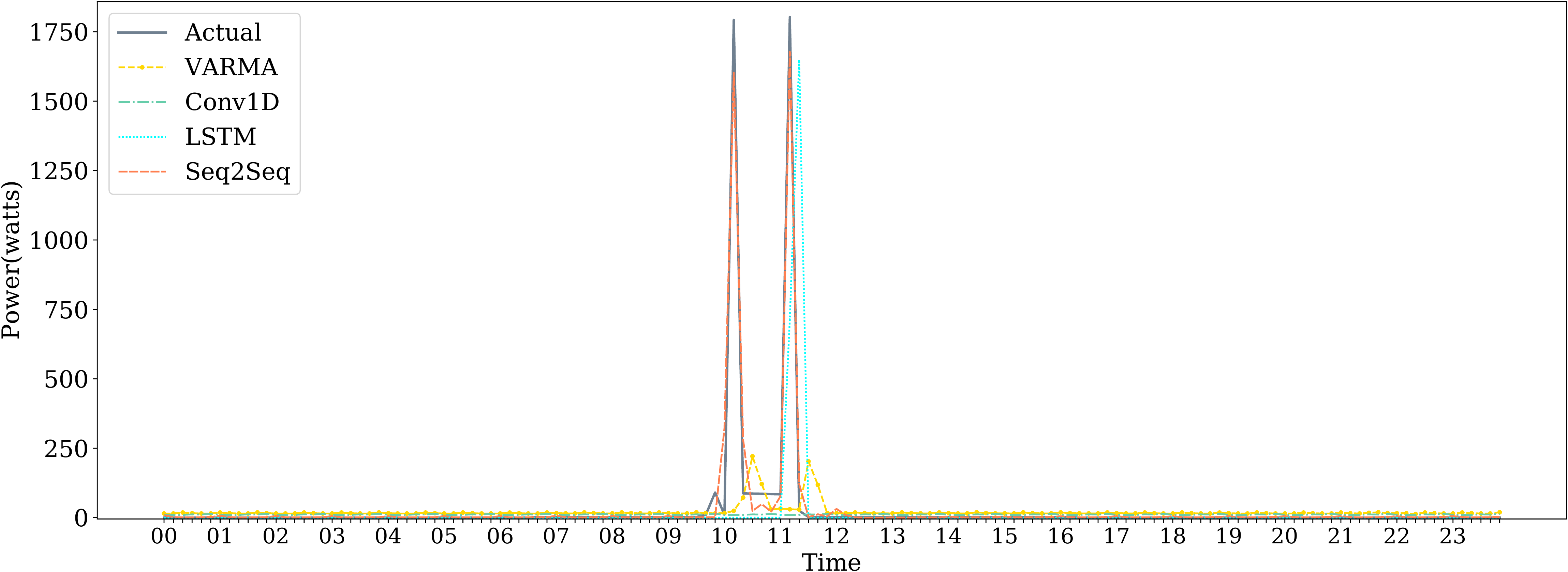}
        \subcaption{Dish Washer}
        \label{fig:dishwasher}
    \end{subfigure}
    
    \begin{subfigure}{0.7\textwidth}
        \includegraphics[width=\linewidth]{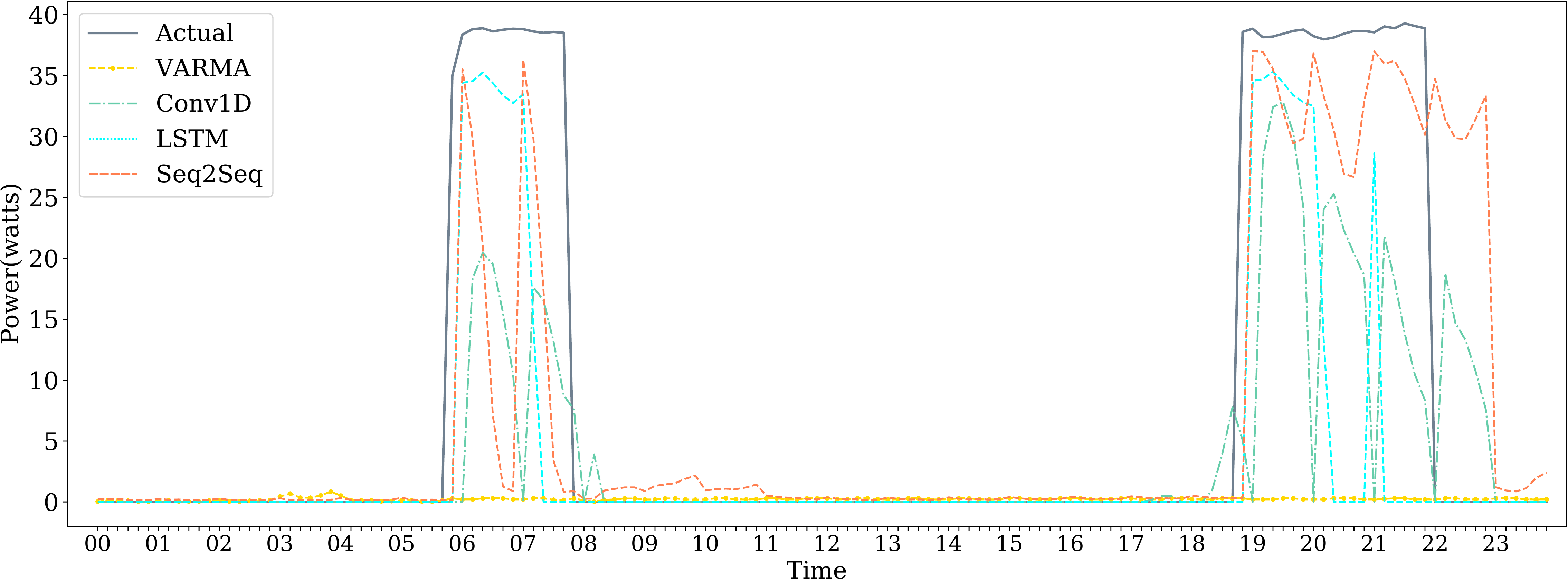}
        \subcaption{Lamp}
        \label{fig:lamp}
    \end{subfigure}
    
    \begin{subfigure}{0.7\textwidth}
        \includegraphics[width=\linewidth]{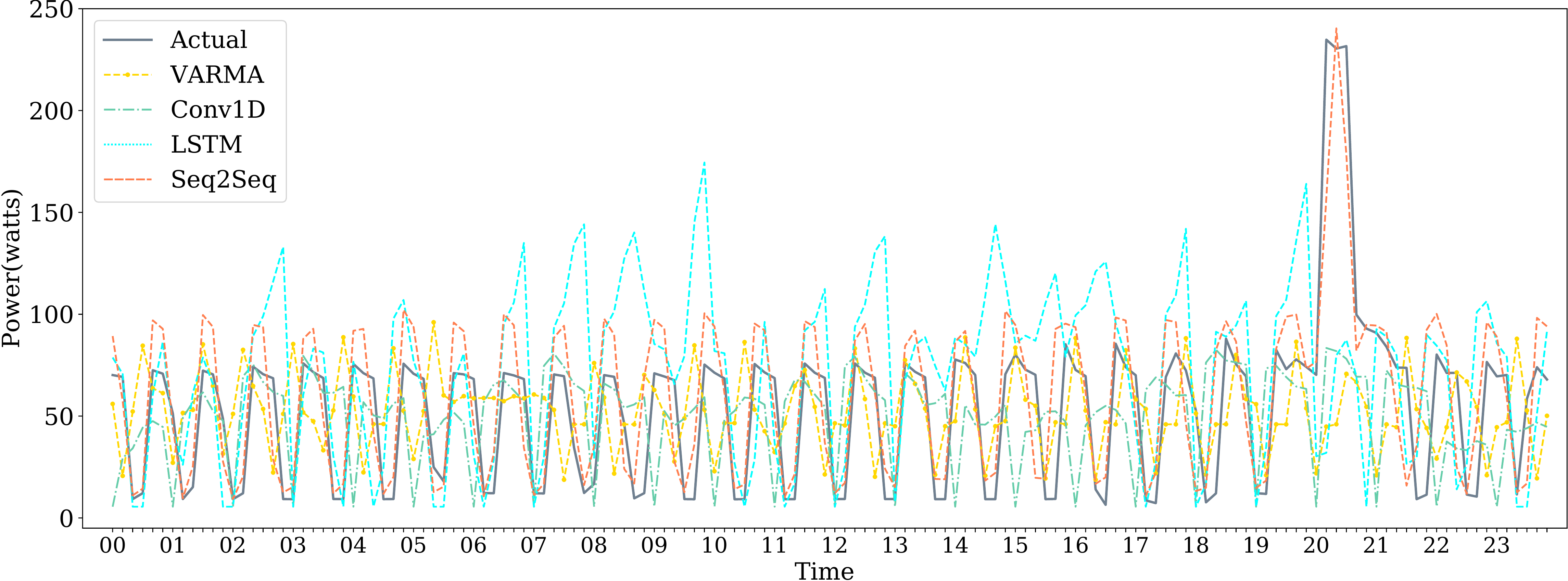}
        \subcaption{Fridge}
        \label{fig:fridge}
    \end{subfigure}
    
    \begin{subfigure}{0.7\textwidth}
        \includegraphics[width=\linewidth]{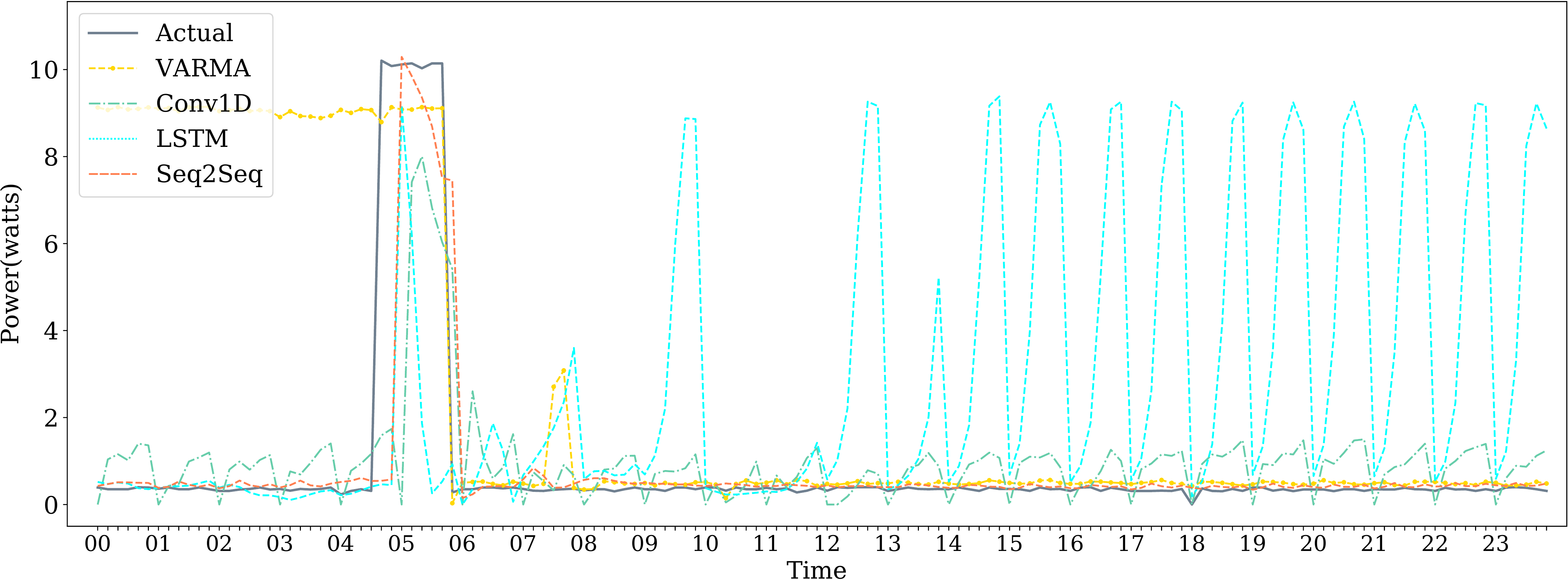}
        \subcaption{Radio}
        \label{fig:radio}
    \end{subfigure}
    
    
\end{figure*}

\begin{figure*}\ContinuedFloat
    \centering
    \begin{subfigure}{0.7\textwidth}
        \includegraphics[width=\linewidth]{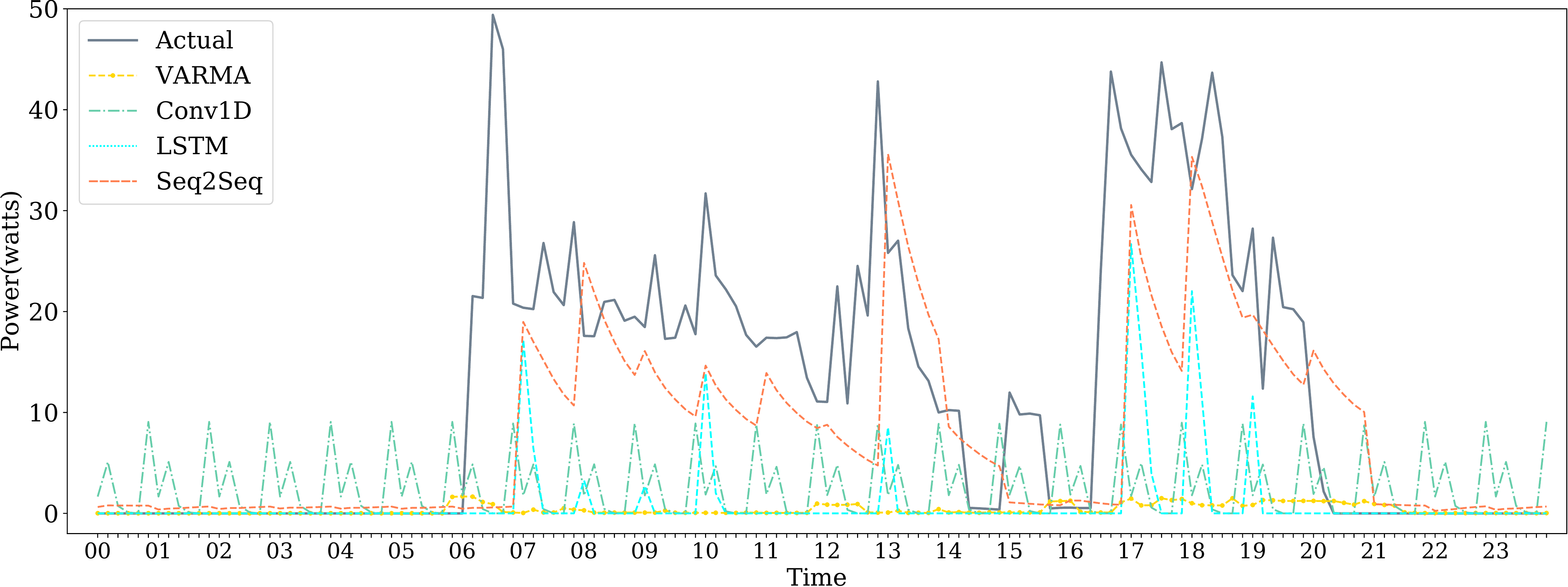}
        \subcaption{Laptop}
        \label{fig:laptop}
    \end{subfigure}
    
    \begin{subfigure}{0.7\textwidth}
        \includegraphics[width=\linewidth]{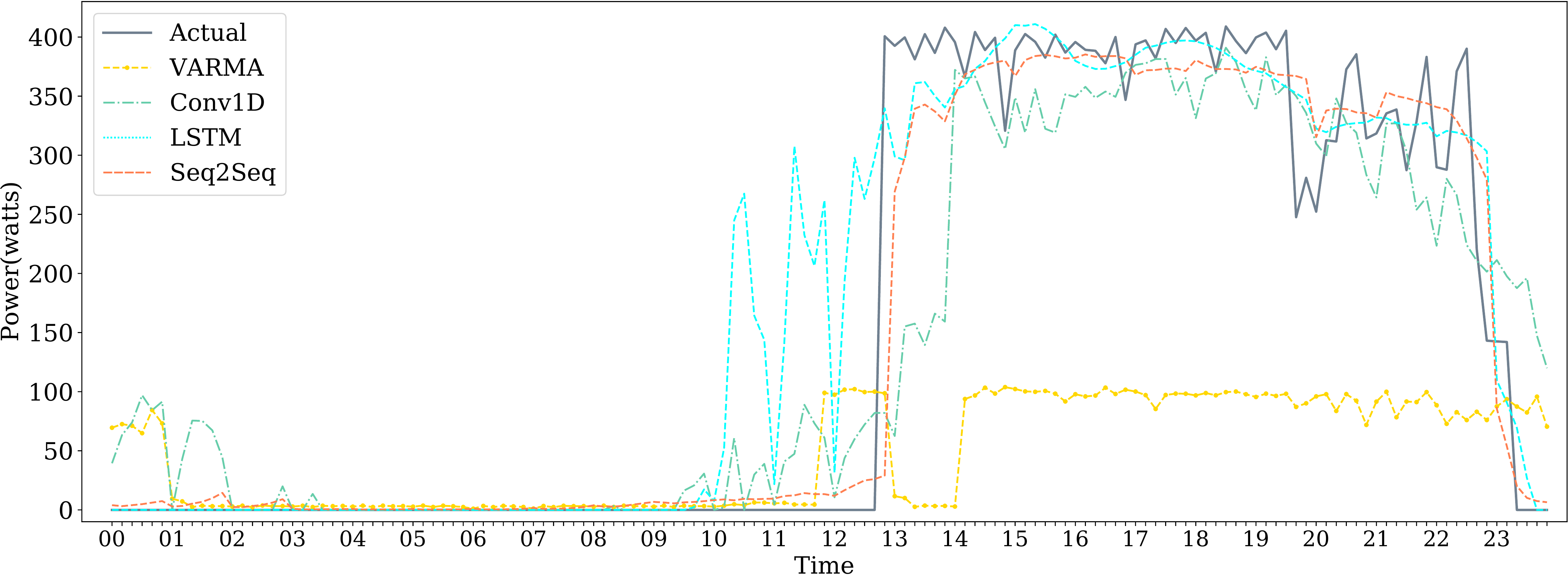}
        \subcaption{TV}
        \label{fig:tv}
    \end{subfigure}
    \caption[]{Energy consumption prediction for 24 hours with a 10-minute resolution.}
    \label{fig:res}
    \vspace{-2mm}
\end{figure*}

\begin{table*}
\caption{Evaluation Results.}
\label{tab:EV-results}
\resizebox{\textwidth}{!}{%
\noindent\begin{tabularx}{\textwidth}{@{}l*{16}{C}c@{}}
\toprule
\multicolumn{16}{c}{\textbf{Building 0}} \\ \cmidrule{2-16}
\multicolumn{1}{c}{\multirow{2}{*}{\textbf{Model}}} &
\multicolumn{3}{c}{Dish Washer} & \multicolumn{3}{c}{Lamp} & \multicolumn{3}{c}{Fridge} & \multicolumn{3}{c}{Radio} & \multicolumn{3}{c}{Washing Machine} \\ \cmidrule(l){2-4} \cmidrule(l){5-7} \cmidrule(l){8-10} \cmidrule(l){11-13} \cmidrule(l){14-16}
\multicolumn{1}{p{0.02\textwidth}}{} & \multicolumn{1}{p{0.02\textwidth}}{RMSE} & \multicolumn{1}{p{0.02\textwidth}}{NRMSE} &
\multicolumn{1}{p{0.02\textwidth}}{MAE}&
\multicolumn{1}{p{0.02\textwidth}}{RMSE} & \multicolumn{1}{p{0.02\textwidth}}{NRMSE} & \multicolumn{1}{p{0.02\textwidth}}{MAE}&
\multicolumn{1}{p{0.02\textwidth}}{RMSE} & \multicolumn{1}{p{0.02\textwidth}}{NRMSE} & \multicolumn{1}{p{0.02\textwidth}}{MAE}&
\multicolumn{1}{p{0.02\textwidth}}{RMSE} & \multicolumn{1}{p{0.02\textwidth}}{NRMSE} & \multicolumn{1}{p{0.02\textwidth}}{MAE}&
\multicolumn{1}{p{0.02\textwidth}}{RMSE} & \multicolumn{1}{p{0.02\textwidth}}{NRMSE} &
\multicolumn{1}{p{0.02\textwidth}}{MAE}\\ \midrule
VARMA & 229.237 & 0.111 & 53.722 & 18.452 & 0.450 & 9.456 & 44.993 & 0.179 & 34.197 & 4.191 & 0.408 & 2.056 & 236.026 & 0.112 & 64.035 \\
Conv1D & 198.400 & 0.096 & 39.421 & 9.837 & 0.240 & 3.842 & 46.133 & 0.183 & 34.073 & 2.395 & 0.233 & 1.156 & \textbf{214.037} & \textbf{0.102} & 51.723 \\
LSTM & 493.948 & 0.239 & 168.236 & 11.632 & 0.284 & 3.942 & 46.601 & 0.185 & 33.218 & 3.513 & 0.342 & 1.805 & 246.735 & 0.117 & \textbf{46.816} \\
Seq2Seq & \textbf{189.799} & \textbf{0.092} & \textbf{37.394} & \textbf{9.185} & \textbf{0.224} & \textbf{3.623} & \textbf{43.748} & \textbf{0.174} & \textbf{30.587} & \textbf{2.314} & \textbf{0.225} & \textbf{0.954} & 215.298 & \textbf{0.102} & 48.373 \\
\addlinespace

\multicolumn{16}{c}{\textbf{Building 1}} \\ \cmidrule{2-16}
\multicolumn{1}{c}{\multirow{2}{*}{}} &
\multicolumn{3}{c}{Dish Washer} & \multicolumn{3}{c}{Lamp} & \multicolumn{3}{c}{Fridge} & \multicolumn{3}{c}{Radio} & \multicolumn{3}{c}{Washing Machine} \\ \cmidrule(l){2-4} \cmidrule(l){5-7} \cmidrule(l){8-10} \cmidrule(l){11-13} \cmidrule(l){14-16}
\multicolumn{1}{p{0.02\textwidth}}{} & \multicolumn{1}{p{0.02\textwidth}}{RMSE} & \multicolumn{1}{p{0.02\textwidth}}{NRMSE} & \multicolumn{1}{p{0.02\textwidth}}{MAE}&
\multicolumn{1}{p{0.02\textwidth}}{RMSE} & \multicolumn{1}{p{0.02\textwidth}}{NRMSE} & \multicolumn{1}{p{0.02\textwidth}}{MAE}&
\multicolumn{1}{p{0.02\textwidth}}{RMSE} & \multicolumn{1}{p{0.02\textwidth}}{NRMSE} & \multicolumn{1}{p{0.02\textwidth}}{MAE}&
\multicolumn{1}{p{0.02\textwidth}}{RMSE} & \multicolumn{1}{p{0.02\textwidth}}{NRMSE} & \multicolumn{1}{p{0.02\textwidth}}{MAE}&
\multicolumn{1}{p{0.02\textwidth}}{RMSE} & \multicolumn{1}{p{0.02\textwidth}}{NRMSE} & \multicolumn{1}{p{0.02\textwidth}}{MAE}\\ \midrule
VARMA & 166.513 & 0.091 & 36.364 & 17.736 & 0.168 & 6.140 & 39.480 & 0.330 & 31.445 & 4.615 & 0.246 & 1.766 & 140.496 & 0.069 & 23.545 \\
Conv1D & 164.924 & 0.090 & 27.609 & 14.757 & 0.140 & 5.298 & 25.534 & 0.213 & 15.541 & 2.860 & 0.152 & 0.867 & 129.324 & 0.063 & 11.285 \\
LSTM & 154.280 & 0.084 & 18.174 & \textbf{11.855} & \textbf{0.112} & \textbf{2.875} & 36.070 & 0.301 & 20.999 & \textbf{2.211} & \textbf{0.118} & \textbf{0.351} & 120.479 & 0.059 & 33.571 \\
Seq2Seq & \textbf{119.308} & \textbf{0.065} & \textbf{13.782} & 12.803 & 0.121 & 3.420 & \textbf{23.346} & \textbf{0.195} & \textbf{10.920} & \textbf{2.314} & \textbf{0.123} & 0.420 & \textbf{119.215} & \textbf{0.058} & \textbf{11.960} \\
\addlinespace

\multicolumn{16}{c}{\textbf{Building 2}} \\ \cmidrule{2-16}
\multicolumn{1}{c}{\multirow{2}{*}{}} &
\multicolumn{3}{c}{Dish Washer} & \multicolumn{3}{c}{TV} & \multicolumn{3}{c}{Network attached storage} & \multicolumn{3}{c}{Laptop} & \multicolumn{3}{c}{Washing Machine} \\ \cmidrule(l){2-4} \cmidrule(l){5-7} \cmidrule(l){8-10} \cmidrule(l){11-13} \cmidrule(l){14-16}
\multicolumn{1}{p{0.02\textwidth}}{} & \multicolumn{1}{p{0.02\textwidth}}{RMSE} & \multicolumn{1}{p{0.02\textwidth}}{NRMSE} &
\multicolumn{1}{p{0.02\textwidth}}{MAE}&
\multicolumn{1}{p{0.02\textwidth}}{RMSE} & \multicolumn{1}{p{0.02\textwidth}}{NRMSE} & 
\multicolumn{1}{p{0.02\textwidth}}{MAE}&
\multicolumn{1}{p{0.02\textwidth}}{RMSE} & \multicolumn{1}{p{0.02\textwidth}}{NRMSE} & 
\multicolumn{1}{p{0.02\textwidth}}{MAE}&
\multicolumn{1}{p{0.02\textwidth}}{RMSE} & \multicolumn{1}{p{0.02\textwidth}}{NRMSE} & 
\multicolumn{1}{p{0.02\textwidth}}{MAE}&
\multicolumn{1}{p{0.02\textwidth}}{RMSE} & \multicolumn{1}{p{0.02\textwidth}}{NRMSE}  &
\multicolumn{1}{p{0.02\textwidth}}{MAE} \\ \midrule
VARMA & 294.618 & 0.135 & 86.852 & 203.580 & 0.478 & 141.565 & 79.463 & 0.550 & 69.964 & 16.122 & 0.030 & 3.785 & 138.297 & 0.112 & \textbf{23.292} \\
Conv1D & 294.417 & 0.134 & 73.202 & 116.982 & 0.275 & 68.136 & 10.406 & 0.072 & 3.808 & 16.236 & 0.030 & 6.329 & \textbf{137.647} & \textbf{0.111} & 43.017 \\
LSTM & 299.399 & 0.137 & \textbf{63.336} & 124.385 & 0.292 & 61.287 & \textbf{7.906} & \textbf{0.055} & \textbf{2.418} & 15.800 & 0.029 & 3.364 & 177.153 & 0.143 & 98.461 \\
Seq2Seq & \textbf{256.000} & \textbf{0.117} & 69.832 & \textbf{113.047} & \textbf{0.266} & \textbf{54.230} & 9.307 & 0.064 & 2.450 & \textbf{14.537} & \textbf{0.027} & \textbf{3.061} & 142.829 & 0.115 & 34.219 \\

\multicolumn{16}{c}{\textbf{Building 3}} \\ \cmidrule{2-16}
\multicolumn{1}{c}{\multirow{2}{*}{}} &
\multicolumn{3}{c}{Dish Washer} & \multicolumn{3}{c}{TV} & \multicolumn{3}{c}{Fridge} & \multicolumn{3}{c}{Computer} & \multicolumn{3}{c}{Washing Machine} \\ \cmidrule(l){2-4} \cmidrule(l){5-7} \cmidrule(l){8-10} \cmidrule(l){11-13} \cmidrule(l){14-16}
\multicolumn{1}{p{0.02\textwidth}}{} & \multicolumn{1}{p{0.02\textwidth}}{RMSE} & \multicolumn{1}{p{0.02\textwidth}}{NRMSE} &
\multicolumn{1}{p{0.02\textwidth}}{MAE}&
\multicolumn{1}{p{0.02\textwidth}}{RMSE} & \multicolumn{1}{p{0.02\textwidth}}{NRMSE} & 
\multicolumn{1}{p{0.02\textwidth}}{MAE}&
\multicolumn{1}{p{0.02\textwidth}}{RMSE} & \multicolumn{1}{p{0.02\textwidth}}{NRMSE} & 
\multicolumn{1}{p{0.02\textwidth}}{MAE}&
\multicolumn{1}{p{0.02\textwidth}}{RMSE} & \multicolumn{1}{p{0.02\textwidth}}{NRMSE} & 
\multicolumn{1}{p{0.02\textwidth}}{MAE}&
\multicolumn{1}{p{0.02\textwidth}}{RMSE} & \multicolumn{1}{p{0.02\textwidth}}{NRMSE} &
\multicolumn{1}{p{0.02\textwidth}}{MAE} \\ \midrule
VARMA & 157.059 & 0.088 & 29.454 & 34.862 & 0.363 & 14.512 & 70.842 & 0.558 & 49.705 & 38.798 & 0.433 & 25.994 & 189.403 & 0.095 & 39.776 \\
Conv1D & 152.446 & 0.085 & 16.818 & 22.188 & 0.231 & 8.625 & 41.075 & 0.323 & 24.815 & 15.416 & 0.172 & 8.242 & 165.717 & 0.083 & \textbf{20.852} \\
LSTM & 135.093 & 0.076 & \textbf{12.546} & \textbf{18.148} & \textbf{0.189} & \textbf{4.666} & 46.150 & 0.363 & 25.728 & \textbf{14.639} & \textbf{0.164} & \textbf{5.947} & 187.217 & 0.094 & 40.206 \\
Seq2Seq & \textbf{131.433} & \textbf{0.073} & 18.913 & 21.599 & 0.225 & 7.167 & \textbf{32.821} & \textbf{0.258} & \textbf{17.385} & 15.813 & 0.177 & 7.124 & \textbf{164.082} & \textbf{0.082} & 31.318 \\

\bottomrule
\end{tabularx}%
}
\end{table*}

\section{Conclusion}
\label{Conclusion}
In this paper, we proposed an appliance-level load forecasting model for residential homes. Our proposed LSTM-based Sequence to Sequence learning model forecasts energy consumption for smart home appliances for the hour ahead with 10-minute resolution using historical data of the past day. We evaluated the performance of our model compared with the VARMA model, Dilated 1D Convolution, and LSTM model on a publicly available dataset that contains historical data for multiple buildings. The consistent outperforming results demonstrate that the LSTM-based Sequence to Sequence learning model does not depend on the data and can be used for any household. As future work, we intend to combine optimization algorithms with prediction results to provide energy-efficient scheduling for a smart home.


\begin{thebibliography}{10}
\providecommand{\url}[1]{#1}
\csname url@samestyle\endcsname
\providecommand{\newblock}{\relax}
\providecommand{\bibinfo}[2]{#2}
\providecommand{\BIBentrySTDinterwordspacing}{\spaceskip=0pt\relax}
\providecommand{\BIBentryALTinterwordstretchfactor}{4}
\providecommand{\BIBentryALTinterwordspacing}{\spaceskip=\fontdimen2\font plus
\BIBentryALTinterwordstretchfactor\fontdimen3\font minus
  \fontdimen4\font\relax}
\providecommand{\BIBforeignlanguage}[2]{{%
\expandafter\ifx\csname l@#1\endcsname\relax
\typeout{** WARNING: IEEEtran.bst: No hyphenation pattern has been}%
\typeout{** loaded for the language `#1'. Using the pattern for}%
\typeout{** the default language instead.}%
\else
\language=\csname l@#1\endcsname
\fi
#2}}
\providecommand{\BIBdecl}{\relax}
\BIBdecl

\bibitem{Zenginis-2018-TII}
I.~{Zenginis}, J.~S. {Vardakas}, J.~{Abadal}, C.~{Echave}, M.~M. {Gűell}, and
  C.~{Verikoukis}, ``Optimal power equipment sizing and management for
  cooperative buildings in microgrids,'' \emph{IEEE Transactions on Industrial
  Informatics}, vol.~15, no.~1, pp. 158--172, 2019.

\bibitem{b8}
W.~Kong, Z.~Dong, Y.~Jia, D.~Hill, Y.~Xu, and Y.~Zhang, ``Short-term
  residential load forecasting based on {LSTM} recurrent neural network,''
  \emph{IEEE Trans. on Smart Grid}, vol.~10, no.~1, pp. 841--85, Jan. 2019.

\bibitem{b5}
H.~Shi, M.~Xu, and R.~Li, ``Deep learning for household load forecasting-a
  novel pooling deep {RNN},'' \emph{IEEE Trans. on Smart Grid}, vol.~9, no.~5,
  pp. 5271--5280, Sep. 2018.

\bibitem{b6}
W.~Kong, Z.~Dong, B.~Wang, J.~Zhao, and J.~Huang, ``A practical solution for
  non-intrusive type {II} load monitoring based on deep learning and
  post-processing,'' \emph{IEEE Trans. on Smart Grid}, vol.~11, no.~1, p.
  148–160, Jan. 2020.

\bibitem{b12}
K.~Amarasinghe, D.~L. Marino, and M.~Manic, ``Deep neural networks for energy
  load forecasting,'' in \emph{Proc. of IEEE Int. Symp. on Industrial
  Electronics}, Jun. 2017, pp. 1--6.

\bibitem{b13}
A.~Almalaq and G.~Edwards, ``A review of deep learning methods applied on load
  forecasting,'' in \emph{Proc. of IEEE Int. Conf. on Machine Learning and
  Applications}, Dec. 2017, pp. 1--6.

\bibitem{b14}
J.~Zheng, C.~Xu, Z.~Zhang, and X.~Li, ``Electric load forecasting in smart
  grids using long-short-term-memory based recurrent neural network,'' in
  \emph{Proc. of Annual Conf. on Information Sciences and Systems}, Mar. 2017,
  pp. 1--6.

\bibitem{b9}
K.~Yan, W.~Li, Z.~Ji., M.~Qi, and Y.~Du., ``A hybrid {LSTM} neural network for
  energy consumption forecasting of individual households,'' \emph{IEEE
  Access}, vol.~7, pp. 157\,633--157\,642, 2019.

\bibitem{b10}
M.~Sajjad, A.~Khan, A.~Ullah, T.~Hussain, Q.~Ullah, M.~Lee, and S.~Baik, ``A
  novel {CNN-GRU}-based hybrid approach for short-term residential load
  forecasting,'' \emph{IEEE Access}, vol.~8, pp. 143\,759--143\,768, 2020.

\bibitem{Razghandi-2020-Globecom}
M.~{Razghandi} and D.~{Turgut}, ``Residential appliance-level load forecasting
  with deep learning,'' in \emph{Proc. of IEEE GLOBECOM}, Dec. 2020, pp. 1--6.

\bibitem{sutskever2014sequence}
I.~Sutskever, O.~Vinyals, and Q.~V. Le, ``Sequence to sequence learning with
  neural networks,'' in \emph{Proc. of Int. Conf. on Neural Information
  Processing Systems}, 2014, pp. 3104--3112.

\bibitem{b0}
K.~Amasyali and N.~El-Gohary, ``A review of data-driven building energy
  consumption prediction studies,'' \emph{Renewable Sustainable. Energy
  Reviews}, vol.~81, pp. 1192--1205, Jan. 2019.

\bibitem{b1}
C.~Yu, P.~Mirowski, and T.~Ho, ``A sparse coding approach to household
  electricity demand forecasting in smart grids,'' \emph{IEEE Trans. on Smart
  Grid}, vol.~8, no.~2, pp. 738--748, Mar. 2017.

\bibitem{b2}
D.~Wu, B.~Wang, D.~Precup, and B.~Boulet, ``Multiple kernel learning based
  transfer regression for electric load forecasting,'' \emph{IEEE Trans. on
  Smart Grid}, vol.~11, no.~2, pp. 1183--1192, Aug. 2019.

\bibitem{b3}
C.~Feng, M.~Sun, and J.~Zhang, ``Reinforced deterministic and probabilistic
  load forecasting via {Q-learning} dynamic model selection,'' \emph{IEEE
  Trans. on Smart Grid}, vol.~11, no.~2, pp. 1377--1386, Aug. 2019.

\bibitem{b4}
Y.~Yang, W.~Li, T.~Gulliver, and S.~Li, ``Bayesian deep learning-based
  probabilistic load forecasting in smart grids,'' \emph{IEEE Trans. on
  Industrial Informatics}, vol.~16, no.~7, pp. 4703--4713, Jul. 2020.

\bibitem{b7}
N.~Linh and P.~Arboleya, ``Deep learning application to non-intrusive load
  monitoring,'' in \emph{Proc. of IEEE Milan PowerTech}, Jun. 2019, pp. 1--6.

\bibitem{monacchi2014greend}
A.~Monacchi, D.~Egarter, W.~Elmenreich, S.~D'Alessandro, and A.~M. Tonello,
  ``{GREEND:} an energy consumption dataset of households in {I}taly and
  {A}ustria,'' in \emph{Proc. of SmartGridComm}, 2014, pp. 511--516.

\bibitem{oord2016wavenet}
A.~van~den Oord, S.~Dieleman, H.~Zen, K.~Simonyan, O.~Vinyals, A.~Graves,
  N.~Kalchbrenner, A.~Senior, and K.~Kavukcuoglu, ``Wavenet: A generative model
  for raw audio,'' 2016.

\end{thebibliography}


\vspace{0.1in}
\end{document}